\documentclass[aps,prb,twocolumn,twoside,superscriptaddress]{revtex4}
\usepackage[version=3]{mhchem} 
\usepackage{amsmath}
\usepackage{amssymb}
\usepackage{amsthm}
\usepackage{amsfonts}
\usepackage{listings}
\usepackage{longtable}
\usepackage{enumerate}
\usepackage{latexsym}
\usepackage{color}
\usepackage{setspace} 
\usepackage{blindtext}
\usepackage{dsfont}
\usepackage{mathrsfs}
\usepackage{array,etoolbox}
\usepackage{bm}
\usepackage{hyperref}
\usepackage{graphicx}
\usepackage{subfigure}
\usepackage{psfrag}
\usepackage{changes}
\usepackage{ulem}

\begin{document}

\newcommand{\TODO}[1]{\textcolor{red}{#1}}
\newcommand{\super}[1]{\ensuremath{^{\mathrm{#1}}}}

\title{Substrate screening approach for quasiparticle energies of
    two-dimensional interfaces with lattice mismatch}
\setcounter{page}{1}

\author{Chunhao Guo}
\affiliation{%
 Department of Chemistry and Biochemistry, University of California Santa Cruz, Santa Cruz, California, 95064, USA
}%
\author{Junqing Xu}
\affiliation{%
 Department of Chemistry and Biochemistry, University of California Santa Cruz, Santa Cruz, CA, 95064, USA
}%
\author{Dario Rocca}
\affiliation{%
 University of Lorraine, LPCT, UMR 7019, 54506 Vandœuvre-lès-Nancy, France
}%
\author{Yuan Ping}
\email{yuanping@ucsc.edu}
\affiliation{%
 Department of Chemistry and Biochemistry, University of California Santa Cruz, Santa Cruz, CA, 95064, USA
}%
\date{\today}

\begin{abstract}
Two-dimensional (2D) materials are outstanding platforms for exotic physics and emerging applications by forming interfaces. In order to efficiently take into account the substrate screening in the quasiparticle energies of 2D materials, several theoretical methods have been proposed previously; but only applicable to interfaces of two systems' lattice constants with certain integer proportion, which often requires a few percent of strain.   
In this work, we analytically showed the equivalence and distinction among different approximate methods for substrate dielectric matrices. 
 We evaluated the accuracy of these methods, by applying them to calculate quasi-particle energies of hexagonal boron nitride interface systems (heterojunctions and bilayers), and compared with explicit interface calculations. 
 Most importantly, we developed an efficient and accurate interpolation technique for dielectric matrices that made quasiparticle energy calculations possible for arbitrarily mismatched interfaces free of strain, which is extremely important for practical applications. 
\end{abstract}

\maketitle


\section{Introduction}

 Two-dimensional (2D) materials and their interfaces have shown unprecedented rich physics and promising applications in many areas, such as opto-spintronics~\cite{wolf2001spintronics, vzutic2004spintronics}, quantum information~\cite{he2015single, tran2016quantum}, and biomedical research~\cite{wang2015ultrathin, sun2015ultrasmall}.
 New emerging phenomena such as non-conventional superconductivity~\cite{cao2018unconventional,cao2018correlated} or topologically protected states~\cite{song2019all, ahn2019failure, qin2014persistent, qin2019chiral} may be created by stacking 2D layers. Experimentally, growth of 2D materials, achieved through physical epitaxy or chemical vapor deposition (CVD), is typically supported on a substrate~\cite{novoselov20162d}. In general, the electrical and optical properties of 2D materials could be strongly modified by substrate screening. For example, the 2D materials' fundamental electronic gap can be significantly reduced due to the dielectric screening from surrounding layers (substrates) when forming heterointerfaces~\cite{winther2017band,ugeda2014giant}.  Reliable prediction of substrate screening effects from first-principles calculations is critical for
 accurate interpretation of experimental results and guidance of new materials' design. 

Currently, widely-used electronic structure methods such as the HSE06 hybrid functional~\cite{krukau2006influence} may accurately describe a large number of three-dimensional bulk systems, but are inadequate for low dimensional systems such as ultrathin 2D materials because of their highly inhomogeneous dielectric screening.
The Koopman's compliant hybrid functional~\cite{stein2010fundamental, nguyen2018koopmans,WLW2018} or dielectric dependent hybrid functional~\cite{zheng2019dielectric} are necessary for the electronic structure of ultrathin 2D materials, where the fraction of Fock exchange $\alpha$ varies with the number of layers~\cite{smart2018fundamental} and needs to be determined for each individual material and thickness.

  
On the other hand, many-body perturbation theory (MBPT)~\cite{ping2013electronic, hybertsen1987ab, sangalli2019many} can successfully describe the quasiparticle properties of 2D materials such as fundamental electronic gaps, regardless of their thickness and dielectric properties. Generally, one and two-particle excitations, experimentally corresponding to charged excitations (e.g. photoemission) and neutral excitation (e.g. optical absorption), can be accurately obtained by the GW approximation~\cite{hedin1999correlation,ping2013electronic,wu2017first,govoni2015large,pham2013g} and the Bethe–Salpeter equation~\cite{rohlfing2000electron,wu2019dimensionality,ping2012ab,ping2013optical,rocca2012solution, rocca2010ab} (BSE), respectively. However, explicit interface calculations at this level of theory are extremely computationally demanding and not suitable for the rapid evaluation of the effect of different substrates. 

 
Therefore, several approximate methods have been proposed to compute the quasiparticle properties of interfaces at the cost of primitive cell calculations of the subsystems composing the interface~\cite{trolle2017model, yan2011nonlocal, ugeda2014giant}. Typically, for weakly-bonded Van de Waals (vdW) interfaces, the hybridization between layers is relatively weak and the dominant effect of the substrate consists in modifying the dielectric screening of the material of interest~\cite{yan2011nonlocal}. Within the GW approximation, this effect can be described by approximating the dielectric matrix of an interface in terms of contributions from individual subsystems (the material and the substrate), as proposed in several previous studies~\cite{trolle2017model, yan2011nonlocal, ugeda2014giant}.
Despite the reasonable level of accuracy achieved through these methods, the underlying approximations and connections between different methods have not been carefully evaluated. For example, partially neglecting local-field effect of substrate dielectric screening (i.e. removing in-plane and/or out-of-plane off-diagonal elements of dielectric matrices~\cite{ugeda2014giant, bradley2015probing, qiu2017environmental}) has been a common approximation previously, which was not carefully examined before. We will test the applicability of such approximation in different systems, for both in-plane and out-of-plane components of dielectric matrices.

Most importantly, previous methods can not be applied to arbitrarily lattice-mismatched 2D interfaces, namely an integer relation between lattice constants is necessary ($L \cdot N = \Tilde{L} \cdot \Tilde{N}$, where $L$ and $\Tilde{L}$ are the primitive lattice constants of the two systems at interfaces, and $N (\Tilde{N})$ is an integer number). Forcing lattice-matching or the fulfillment of the above relation is typically required for interface calculations. These constraints either limit the choice of interfaces that can be studied or require applying artificial strain that may strongly modify the electronic structure. In this work, we develop a reciprocal-space linear-interpolation method in the entire $\mathbf{q} + \mathbf{G}$ space to approximate interface dielectric matrices of arbitrarily mismatched systems. This approach makes MBPT calculations of general interfaces possible and free of strain. 

In order to demonstrate the accuracy and efficiency of this new methodology, we will consider applications to interfaces involving hexagonal boron nitride (hBN).
This material has a wide band gap in ultraviolet region, with promising applications in deep ultraviolet light-emitting devices~\cite{kubota2007deep} and as a host for spin qubits and single photon emitters~\cite{FengPRB2019} in quantum information technologies~\cite{awschalom2013quantum, wu2017first}. As ultrathin hBN is mostly supported on substrates in experimental measurements, it is critical to accurately predict the effect of substrates on electronic structure of hBN. This is also important for evaluations of defect properties in 2D materials supported by substrates~\cite{abidi2019selective,wang2019substrate}. We will use hBN with SnS$_2$ substrates and bilayer hBN in two conformations as prototypical examples for our methodology validation in this study. 

For the rest of the paper, we  
first analytically derived the connection among different approximations of dielectric matrices with substrate screening~\cite{yan2011nonlocal,ugeda2014giant,qiu2017environmental, xuan2019quasiparticle,liu2019accelerating}. 
We then performed the separate GW calculations for subsystems from interfaces with several approximate approaches
 to construct the interface polarizability, and compared  results with explicit interfaces in order to evaluate the accuracy of these methods. Next we examine the importance of off-diagonal elements of polarizability in substrate screenings in various 2D interface systems. Finally, we introduced our linear-interpolation technique, benchmarked it and showed the quasiparticle energies obtained by this technique for arbitrarily lattice-mismatched 2D interfaces. 

\section{Methodology} \label{method}

In this section, we will discuss the different methods and concepts used in this paper, which are summarized in the Table~\ref{tab:table1}.  


\begin{table}[t]

\caption{\label{tab:table1}%
Overview of methodology in this work. 
}
\begin{ruledtabular}
\footnotesize
\begin{tabular}{ l c }
\textrm{\textbf{Methods}}&
\textrm{\textbf{Assumption}}\\
\colrule
$\chi_{\text{eff}}$-sum (Eq.~\ref{eq_sum_chi_eff})& Coulomb interaction between layers \\
$\chi^{GSC}_{\text{eff}}$-sum & Uses interface eigenvalue \\ & in $\chi_{\text{eff}}$-sum \\

$\chi^{FWF}_{\text{eff}}$-sum & Uses interface
eigenvalue \\ & and wavefunctions in GW with $\chi_{\text{eff}}$-sum\\
$\chi_0$-sum (Eq.~\ref{eq_sum_chi_0}) & Coulomb interaction between layers, \\& equiv. to $\chi_{\text{eff}}$-sum at RPA\\
$\chi$-sum (Eq.~\ref{eq_sum_chi})& No interaction between layers\\

\textrm{\textbf{Approximations}}&
\textrm{\textbf{Definition}}\\
\colrule
 $\epsilon^{-1}$-diag & Neglects $\chi^{s}$ off-diagonal elements \\
 $\epsilon$-diag & Neglects $\chi^{s}_0$ off-diagonal elements \\

\textrm{\textbf{Interface structure}}&
\textrm{\textbf{Solution}}\\
\colrule
 Lattice match & Direct summation \\
 Special match & $\mathbf{q} + \mathbf{G}$ mapping\\
 Arbitrary mismatch & $\mathbf{q} + \mathbf{G}$ bilinear interpolation\\

\end{tabular}
\end{ruledtabular}
\end{table}

\subsection{Methods for interface polarizability}

The interactions among quasi-particles within the GW approximation is described by the screened Coulomb potential $W = \epsilon^{-1} v_C$, where $v_C$ is the bare Coulomb interaction and $\epsilon$ is the dielectric matrix. The inverse dielectric matrix is defined by $\epsilon^{-1} = \mathds{1} + v_C \chi$ within the random phase approximation  (RPA)~\cite{hybertsen1987ab}. The reducible polarizability $\chi$ can be obtained from the irreducible polarizability $\chi_0$ (also known as independent-particle polarizability) through the equation $\chi = \chi_0 + \chi_0 v_C \chi$. 

\begin{figure}
    \includegraphics[width=0.45\textwidth]{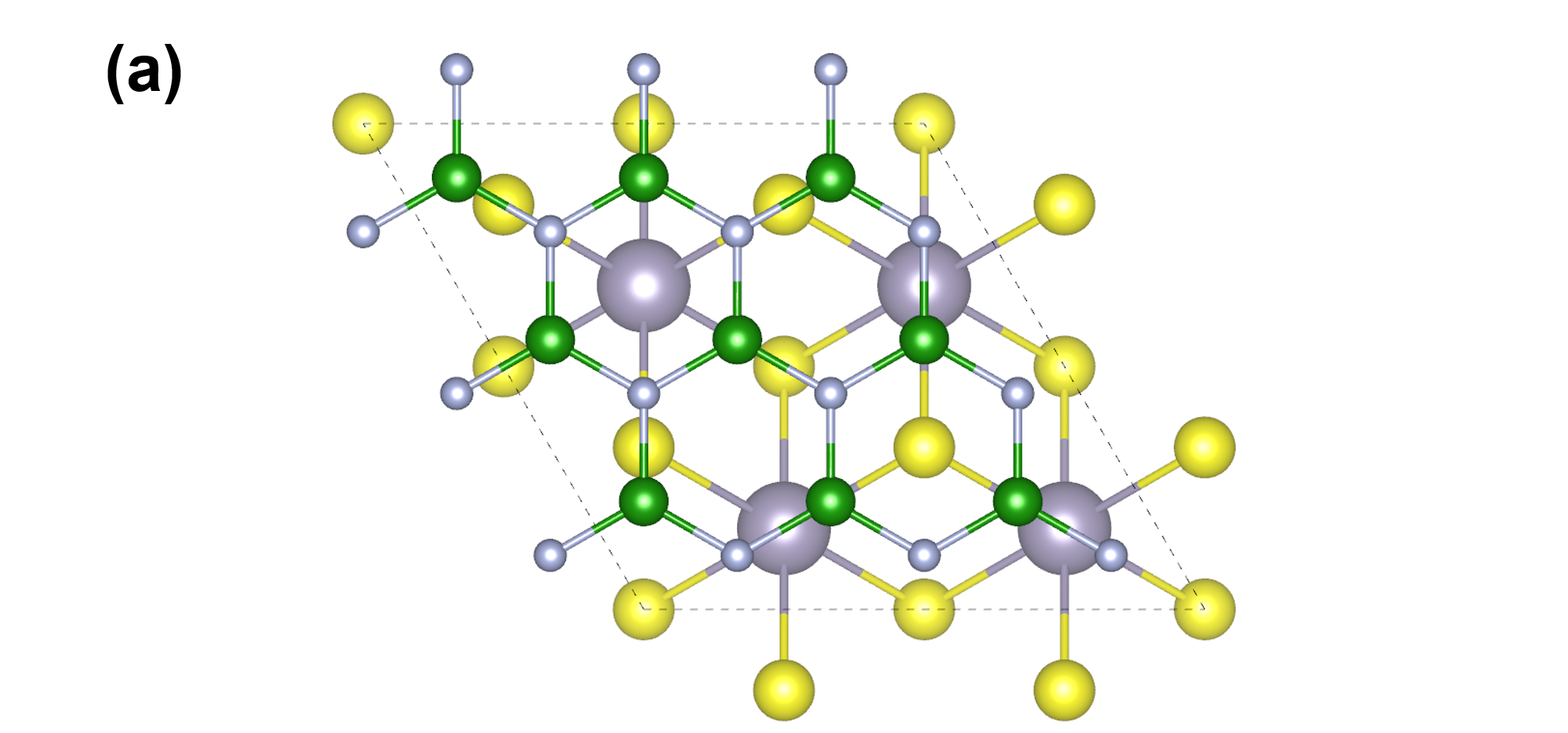}\\
    \includegraphics[width=0.45\textwidth]{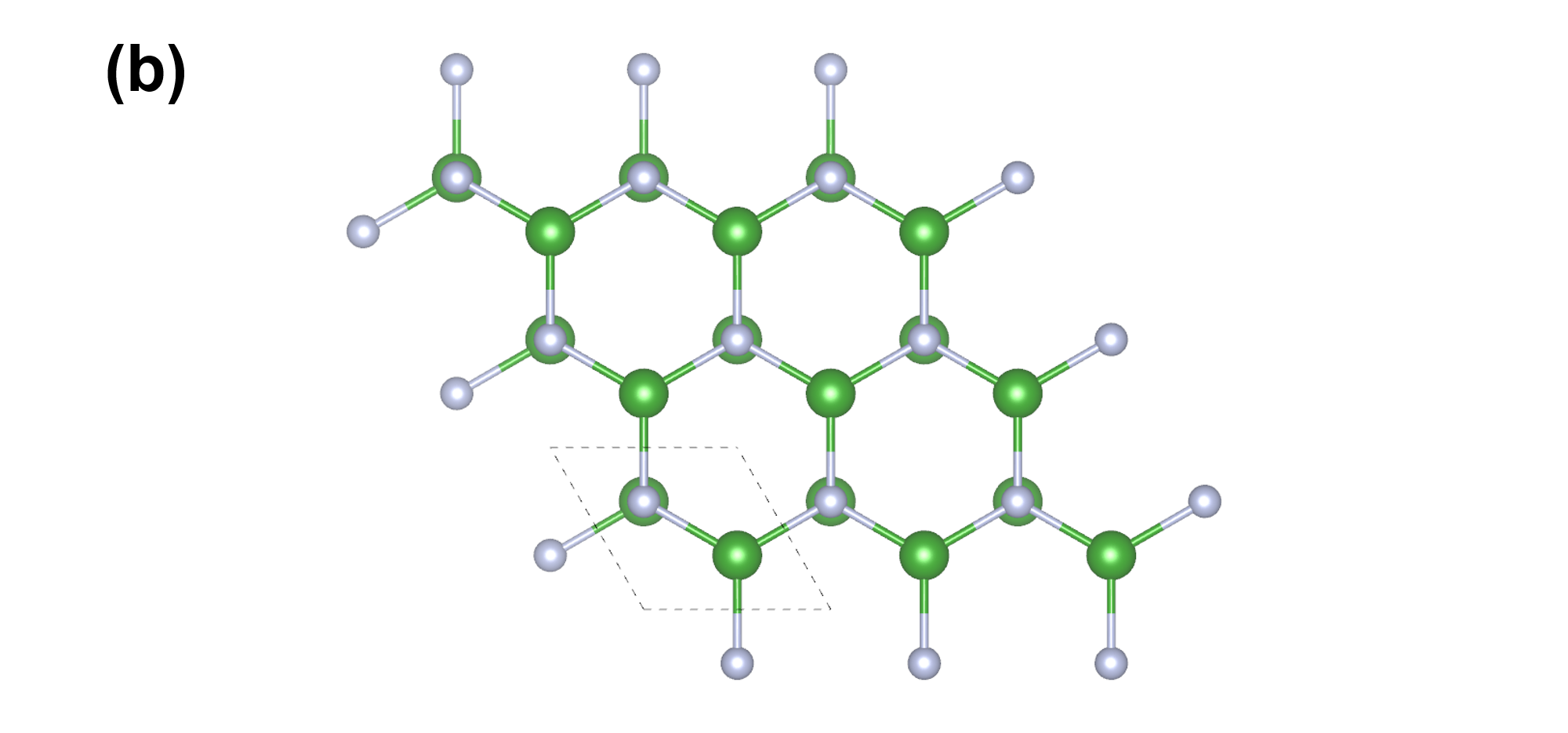}\\
    \includegraphics[width=0.45\textwidth]{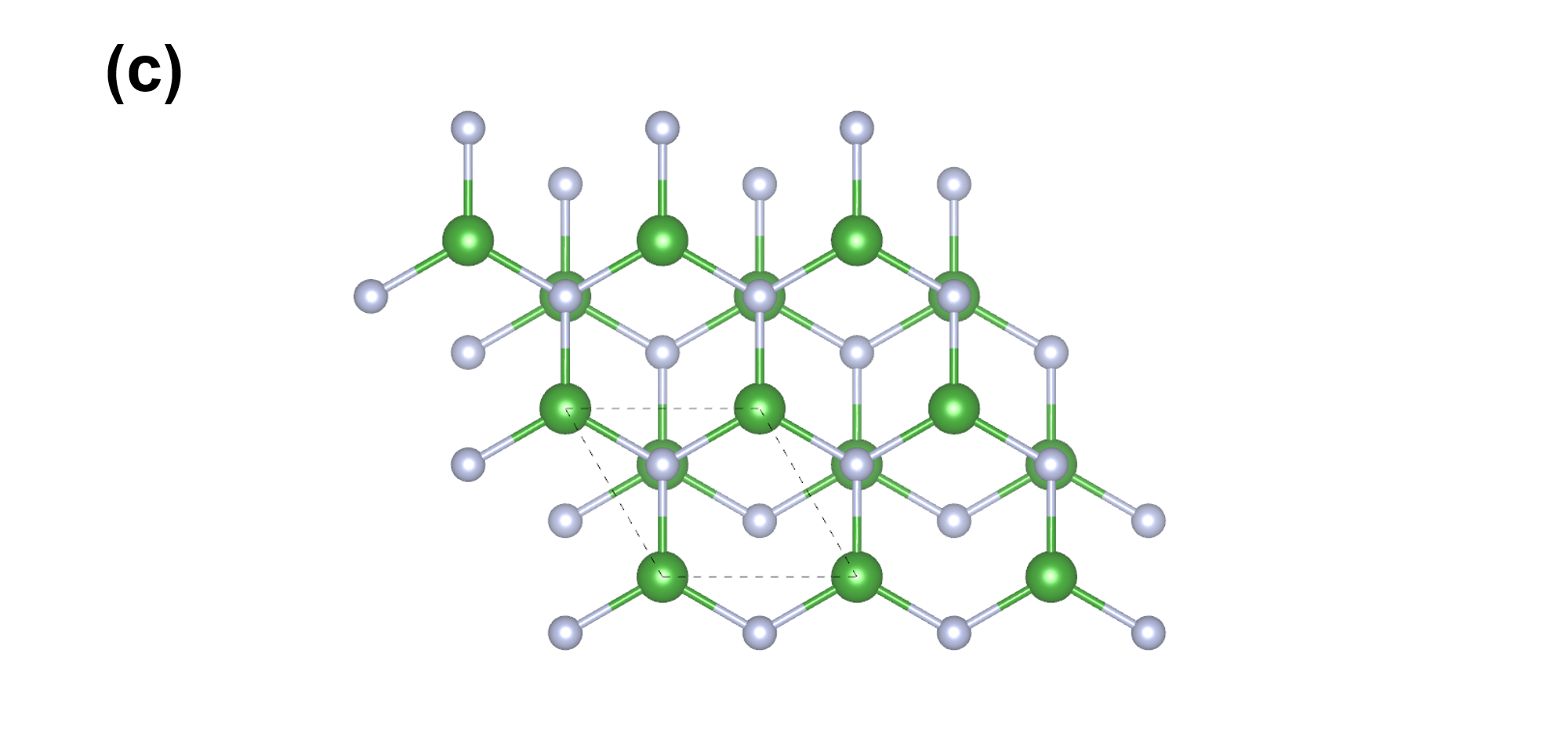}
    \centering
  \caption{\label{atomic} Atomic structure of 2D interfaces a) hBN/SnS$_2$, b) bilayer hBN with AB stacking, c) bilayer hBN with AA$^{\prime}$ stacking.
  The green balls denote boron atoms; the white balls denote N atoms; the yellow balls denote sulfur atoms; the silver balls denote Sn atoms.}
\end{figure}

For the purpose of our discussion we first partition the total (``tot") vdW  interface systems into material (``m") and substrate (``s") subsystems~\cite{yan2011nonlocal}. 
Considering density response of external field, we obtain:
\begin{equation} \label{eq_lin_res}
\begin{split}
\delta n^m =& \chi^m (\delta V_{\text{ext}} + v_C \delta n^s), \\
\delta n^s =& \chi^s (\delta V_{\text{ext}} + v_C \delta n^m),
\end{split}
\end{equation}
where $\delta n$ is the density response, and the reducible polarizability $\chi$ is defined as the density-density response function to an applied potential. 
If we consider the material subsystem (``m") as the probe, the total external potential includes the external applied potential ($\delta V_{\text{ext}}$) and the Coulomb potential from the charge response $\delta n^s$ in the substrate ($v_C \delta n^s$). (We assume the material and substrate are  connected only through interlayer Coulomb interactions, with minimum wavefunction overlap between material and substrates, i.e. interlayer hybridization.)  
Then we define an effective polarizability $\chi_{\text{eff}}$ as a density response function of one subsystem to only the external applied  potential  ($\delta V_{\text{ext}}$), i.e.
$\chi_{\text{eff}}^{m/s} = \delta n^{m/s} / \delta V_{\text{ext}}$. More precisely, $\chi_{\text{eff}}^{m/s}$ can be given in terms of $\chi^{m/s}$ through Eq.~\ref{eq_lin_res}:

\begin{equation} \label{eq_chi_eff}
\begin{split}
&\chi^{m/s}_{\text{eff}} = \frac{\delta n^{m/s}}{\delta V_{\text{ext}}} \\=& ({\mathds{1} - {\chi}^{m/s} v_C {\chi}^{s/m} v_C})^{-1}({\chi}^{m/s} + {\chi}^{m/s} v_C {\chi}^{s/m}).
\end{split}
\end{equation}

When subsystems have negligible interlayer wavefunction overlap (i.e. hybridization), the total density response ($\delta n^{\text{tot}}$) can be written as $\delta n^{\text{tot}} = \delta n^{m} + \delta n^{s}$ and then the total polarizability $\chi^{\text{tot}}$ of entire interface systems is
\begin{equation} \label{eq_sum_chi_eff}
\begin{split}
    \chi^{\text{tot}} =& \frac{\delta n^{\text{tot}}}{\delta V_{\text{ext}}} =  \frac{\delta n^{\text{m}}}{\delta V_{\text{ext}}} +  \frac{\delta n^{\text{s}}}{\delta V_{\text{ext}}} = {\chi}^{m}_{\text{eff}} + {\chi}^{s}_{\text{eff}}.
\end{split}
\end{equation}

In summary, this approach uses the reducible polarizabilty of each subsystem ($\chi^{m/s}$) where the Coulomb potential from the other subsystem is considered part of external potential in Eq.~\ref{eq_lin_res},  to construct the effective reducible polarizability $\chi_{\text{eff}}^{m/s}$ of each subsystem where such potential is excluded from external potential in Eq.~\ref{eq_chi_eff}.
Then we sum up $\chi_{\text{eff}}^{m}$ and $\chi_{\text{eff}}^{s}$ to obtain total reducible polarizability of interface systems $\chi^{\text{tot}}$ in Eq.~\ref{eq_sum_chi_eff}, which will be denoted as ``$\chi_{\text{eff}}$-sum''.

As we noted above, interlayer wavefunction overlap or hybridization effect is not taken into account in the method described above. The hybridization effect can change the eigenvalues and eigenfunctions at the DFT level which then change the Green's function (G) and dielectric matrix (in W) in the GW calculations. Therefore, for systems with strong interlayer hybridization, we can add the hybridization effect step-by-step. We can add  corrections from ground state eigenvalues of interfaces to the $\chi_{\text{eff}}$-sum methods, namely ``$\chi^{GSC}_{\text{eff}}$-sum" method, which partially take into account the effect of interlayer hybridization on eigenvalues at the DFT level. Furthermore, we can also include interface ground state wavefunction (``FWF") and eigenvalues as inputs for Green's function (G), denoted as ``$\chi^{FWF}_{\text{eff}}$-sum" method. This method is close to  GW calculations of an explicit interface except with approximate dielectric matrix by Eq.~\ref{eq_sum_chi_eff}.  

From another perspective, if the interlayer hybridization or wavefunction overlap is negligible (similar to the condition required above for $\delta n^{\text{tot}}$)~\cite{liu2019accelerating, xuan2019quasiparticle}, the total irreducible polarizability $\chi^{tot}_0$ of the interfaces can be expressed approximately as the sum of each subsystem contribution~\cite{bradley2015probing,ugeda2014giant,qiu2017environmental,naik2018substrate,xuan2019quasiparticle,liu2019accelerating}

\begin{equation} \label{eq_sum_chi_0}
\begin{split}
    {\chi}^{\text{tot}}_0 = {\chi}^{\text{m}}_0 + {\chi}^{\text{s}}_0,
\end{split}
\end{equation}
which we denote as ``$\chi_0$-sum" method. To further understand the theoretical connection between different methods, we rewrite Eq.~\ref{eq_lin_res} with ${\chi}_0$
through relation $\chi = \chi_0 + \chi_0 v_C\chi$
as:

\begin{equation} \label{eq_lin_res_tot}
\begin{split}
\delta n^m =& \chi^m_0 (\delta V_{\text{ext}} + v_C \delta n^s + v_C \delta n^m), \\
\delta n^s =& \chi^s_0 (\delta V_{\text{ext}} + v_C \delta n^m + v_C \delta n^s).
\end{split}
\end{equation}
Here $\chi_0$ as the irreducible polarizability is the density response function to total field $\delta V_{tot}$, which includes the applied field and bare Coulomb potential of the total interface system, namely $\delta V_{\text{tot}} \equiv \delta V_{\text{ext}} + v_C \delta n^{\text{tot}}$.
Using the above condition $\delta n^{\text{tot}} = \delta n^{m} + \delta n^{s}$ for the interface, summation of the two equations of subsystems in Eq.~\ref{eq_lin_res_tot} results in $\delta{n^{tot}}=\chi^{\text{tot}}_0 \delta V_{\text{tot}} = (\chi^m_0 + \chi^s_0) \delta V_{\text{tot}}$, which gives Eq.~\ref{eq_sum_chi_0}. This indicate that the $\chi_{\text{eff}}$-sum method and $\chi_0$-sum method are equivalent under RPA. 
However, $\chi_{\text{eff}}$-sum method and $\chi_0$-sum method are not equivalent when the diagonal approximation is applied, i.e. neglecting off-diagonal elements of $\chi$ in the former or $\chi_0$ in the latter, as we will discuss in the Sec. II.B. Therefore we primarily used $\chi_{\text{eff}}$-sum method in this paper.

If we further neglect the interlayer Coulomb interaction, this will set $v_C \delta n^{m/s}$ to zero in Eq.~\ref{eq_lin_res}  and lead
$\chi_{\text{eff}} \to \chi$. This is at the non-interacting limit between two layers, where
\begin{equation} \label{eq_sum_chi}
 {\chi}^{\text{tot}} = {\chi}^{\text{m}} + {\chi}^{\text{s}},
 \end{equation} 
and we name it as ``$\chi$-sum'' method.
In Sec.~\ref{method-compare}, we will compare the quasiparticle energies of interfaces with the above approximated dielectric matrices with explicit interface GW calculations.  

\subsection{Diagonal approximation of dielectric screening}

For simple metals which may be treated as ``jellium", the nearly translational invariance justifies the dielectric matrix $\epsilon$ may be diagonal in reciprocal space~\cite{hybertsen1987ab}. However, semiconductors and insulators have strong in-homogeneity at interaction length scale requires non-zero off-diagonal elements of $\epsilon$~\cite{resta1981dielectric, hybertsen1987ab}. The effect from off-diagonal elements of dielectric matrix $\epsilon$ is often referred to the ``local field effect"~\cite{ceperley1980ground, resta1981dielectric, hybertsen1987ab}.

While the effect of off-diagonal terms in intrinsic dielectric screening has been systematically studied~~\cite{ceperley1980ground, resta1981dielectric, hybertsen1987ab}, the off-diagonal terms' effect from environmental dielectric screening has not been studied in detail. Here we will investigate the off-diagonal effect of environmental dielectric screening through 
two different approaches, i.e. by applying the diagonal approximation of dielectric matrix $\epsilon$ (``$\epsilon$-diag", which directly relates to diagonal approximation of $\chi_0$) or inverse dielectric matrix $\epsilon^{-1}$ (``$\epsilon^{-1}$-diag", which directly relates to diagonal approximation of $\chi$ and $\chi_{\text{eff}}$).

The $\epsilon$-diag approximation has been used for substrate dielectric screening in the past work~\cite{bradley2015probing,ugeda2014giant,qiu2017environmental,naik2018substrate} when applying the $\chi_0$-sum method, specifically, by removing the in-plane off-diagonal components of substrate dielectric matrices.  The $\epsilon^{-1}$-diag approximation has not been employed before, but is more convenient in the $\chi_{\text{eff}}$-sum approach. Since the off-diagonal elements of $\chi_0$ will contribute to the diagonal elements of $\chi$ and $\epsilon^{-1}$ through the matrix inverse operation, 
this is a weaker approaximation than $\epsilon$-diag. We will compare these two approximations considering specific numerical examples in Sec.~\ref{diag-compare}.

\subsection{Reciprocal-space linear-interpolation approach}\label{RLM}

The construction of  interface structure models is often complicated by the problem of lattice matching between two subsystems. One of the main objectives of this work is to propose a general approach that can be applied to subsystems with rather different periodicity and crystal symmetry, and does not require the application of strain to force the lattice matching at the interface.  

\begin{figure}
    \includegraphics[width=0.3\textwidth]{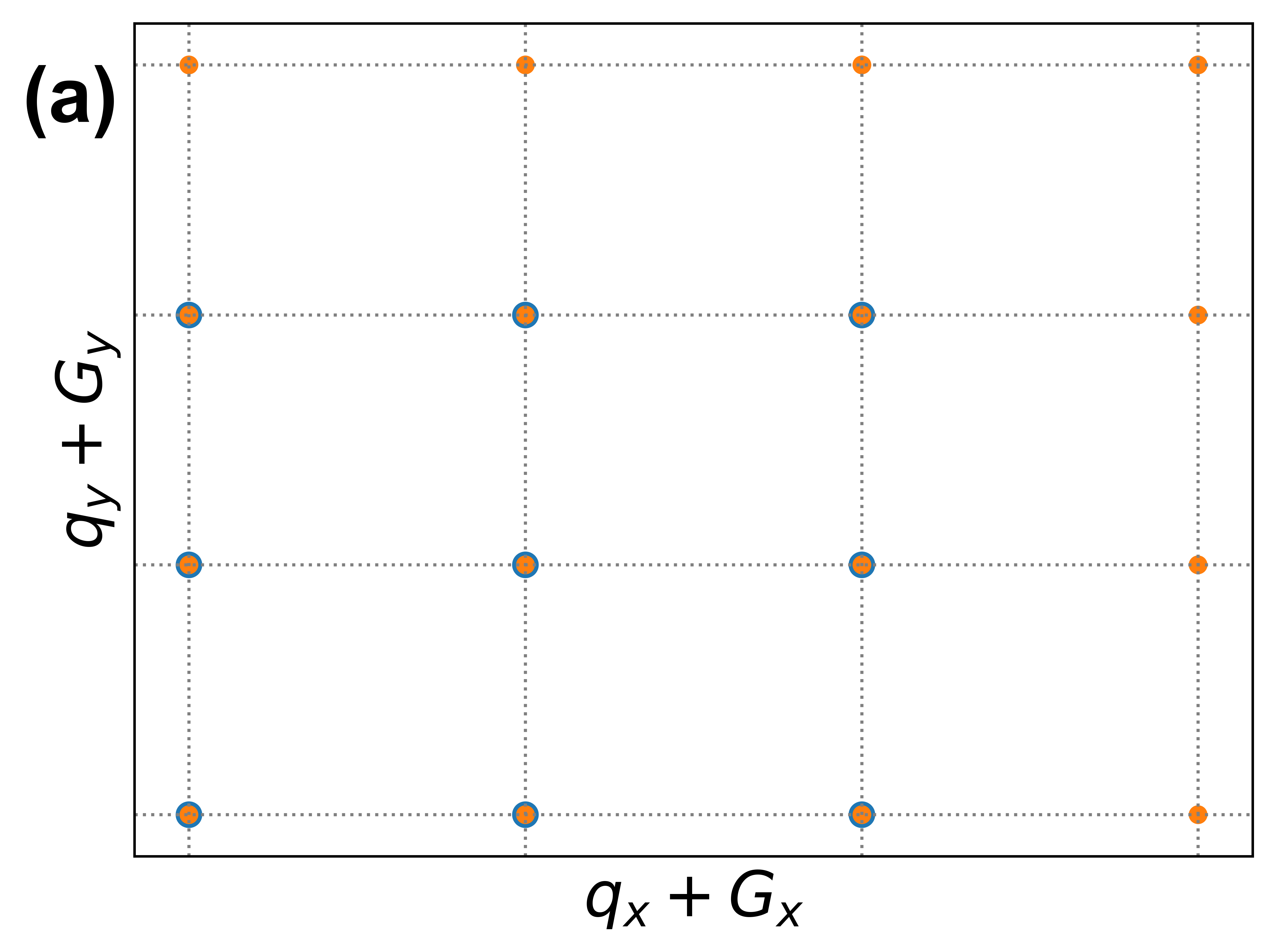}\\
    \includegraphics[width=0.3\textwidth]{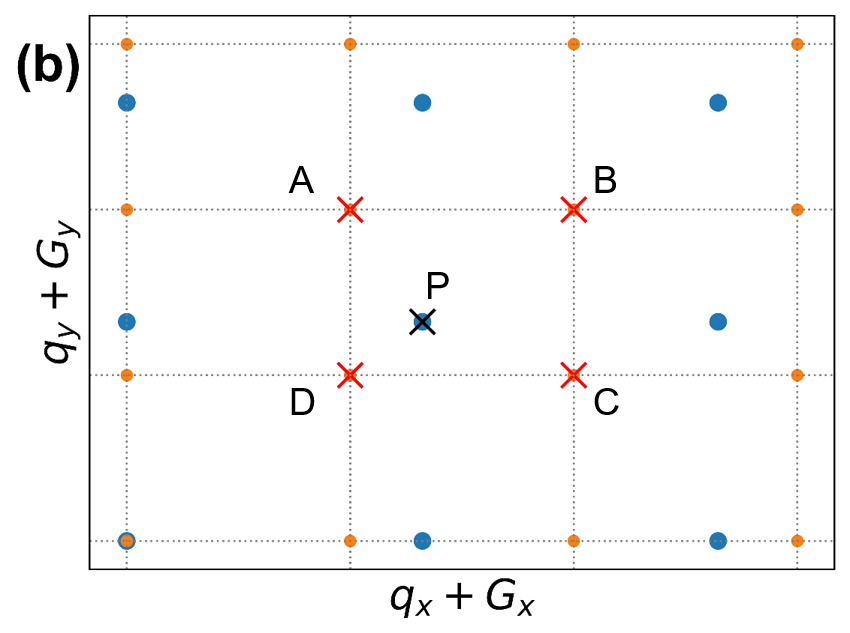}\\
    \includegraphics[width=0.3\textwidth]{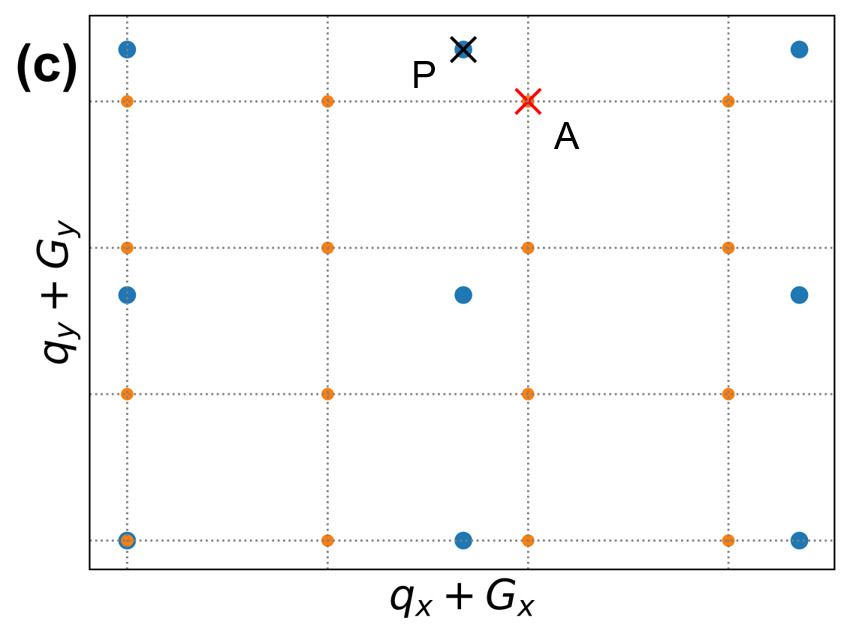}
    \centering
  \caption{\label{interp1} Schematic diagram of a) mapping between computed data points of substrate (orange dots) and computed data points of material (blue dots), where the reciprocal space $\mathbf{q} + \mathbf{G}$ grid from the substrate and $\Tilde{\mathbf{q}} + \Tilde{\mathbf{G}}$ grid from the material are overlapping; b) bilinear interpolation of the black cross point at $P$ (from $\Tilde{\mathbf{q}} + \Tilde{\mathbf{G}}$ grid) from the four nearest data points $A,B,C,D$ (red cross) when $P$ is inside the boundary of the $\mathbf{q} + \mathbf{G}$ grid (orange dots); c) proximal interpolation with the only nearest one point $A$ when the interpolation point is $P$ at the boundary.}
\end{figure}

In general, in order to directly sum the subsystem contributions to obtain the  polarizability (and dielectric matrix) of the full interface, one needs an exact correspondence of the $\mathbf{q} + \mathbf{G}$ vectors between the material and the substrate. This requires finding two integer numbers $N$ and $\Tilde{N}$ such that the lattice constants $L$ (substrate) and $\Tilde{L}$ (material) satisfy the relation $L \cdot N = \Tilde{L} \cdot \Tilde{N}$. If $N$ and $\Tilde{N}$ can be chosen to be reasonably small, calculations can be directly performed for supercells containing $N$ and $\Tilde{N}$ repetitions, although this approach often requires the application of a small percentage of strain. 
However, if the required $N$ or $\Tilde{N}$ are large, several methods have been proposed to make this type of calculations practical~\cite{bradley2015probing,ugeda2014giant,qiu2017environmental,naik2018substrate,xuan2019quasiparticle,liu2019accelerating}. The central idea is to consider unit cells only and to perform a one to one mapping between the reciprocal space $\mathbf{q} + \mathbf{G}$ vectors of the material and substrate~\cite{liu2019accelerating} (see Figure~\ref{interp1}(a)). This approach still requires the relation $L \cdot N = \Tilde{L} \cdot \Tilde{N}$ to be satisfied (possibly by applying a small strain to modify $L$ or $\Tilde{L}$) but avoids supercell calculations. We note that even if one applies the diagonal approximation for $\chi_0$ or $\chi$, the diagonal elements still contain both $\mathbf{q}$ and $\mathbf{G}$ vectors, which requires this relation to be satisfied.
While this is a clear numerical improvement, a large number of $\mathbf{q}$ vectors in the first Brillouin zone might still be required.
Indeed, one needs to sample $\mathbf{q}$ and/or $\Tilde{\mathbf{q}}$ point meshes fine enough to ensure that the number of $\mathbf{q}$ points ($N_{\mathbf{q}}$) satisfy the relation $L \cdot N_{\mathbf{q}} = \Tilde{L} \cdot N_{\Tilde{\mathbf{q}}}$ or equivalently $N_{\mathbf{q}}/N_{\Tilde{\mathbf{q}}} = N/\Tilde{N}$. Accordingly, this approach becomes computationally demanding for large $N$ and/or $\Tilde{N}$. A more serious issue is that this mapping scheme is not possible for interfaces with two systems with very different crystal symmetry, e.g. a hexagonal and a   lattice.

\begin{figure}
    \includegraphics[width=0.45\textwidth]{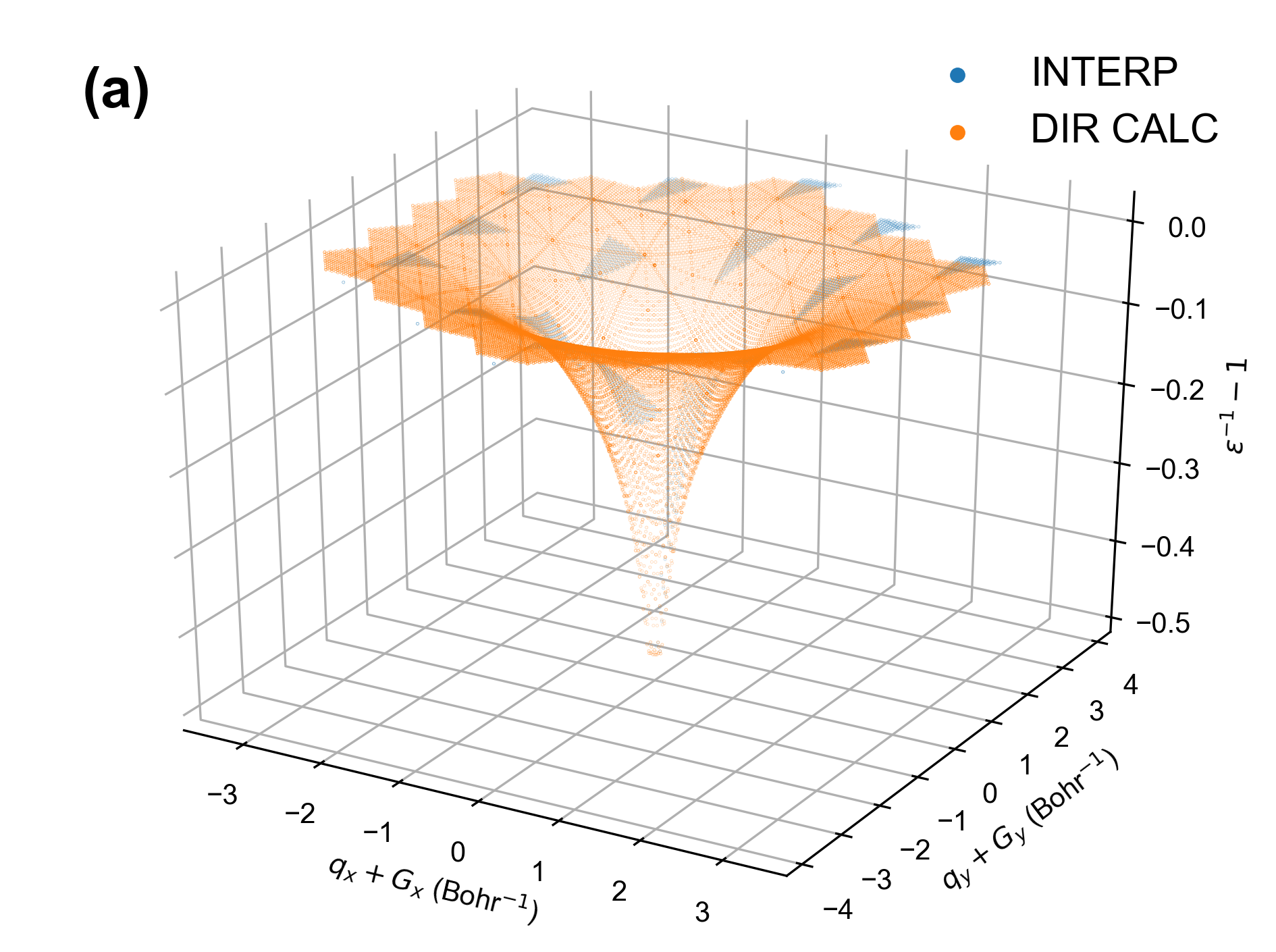}\\
    \includegraphics[width=0.45\textwidth]{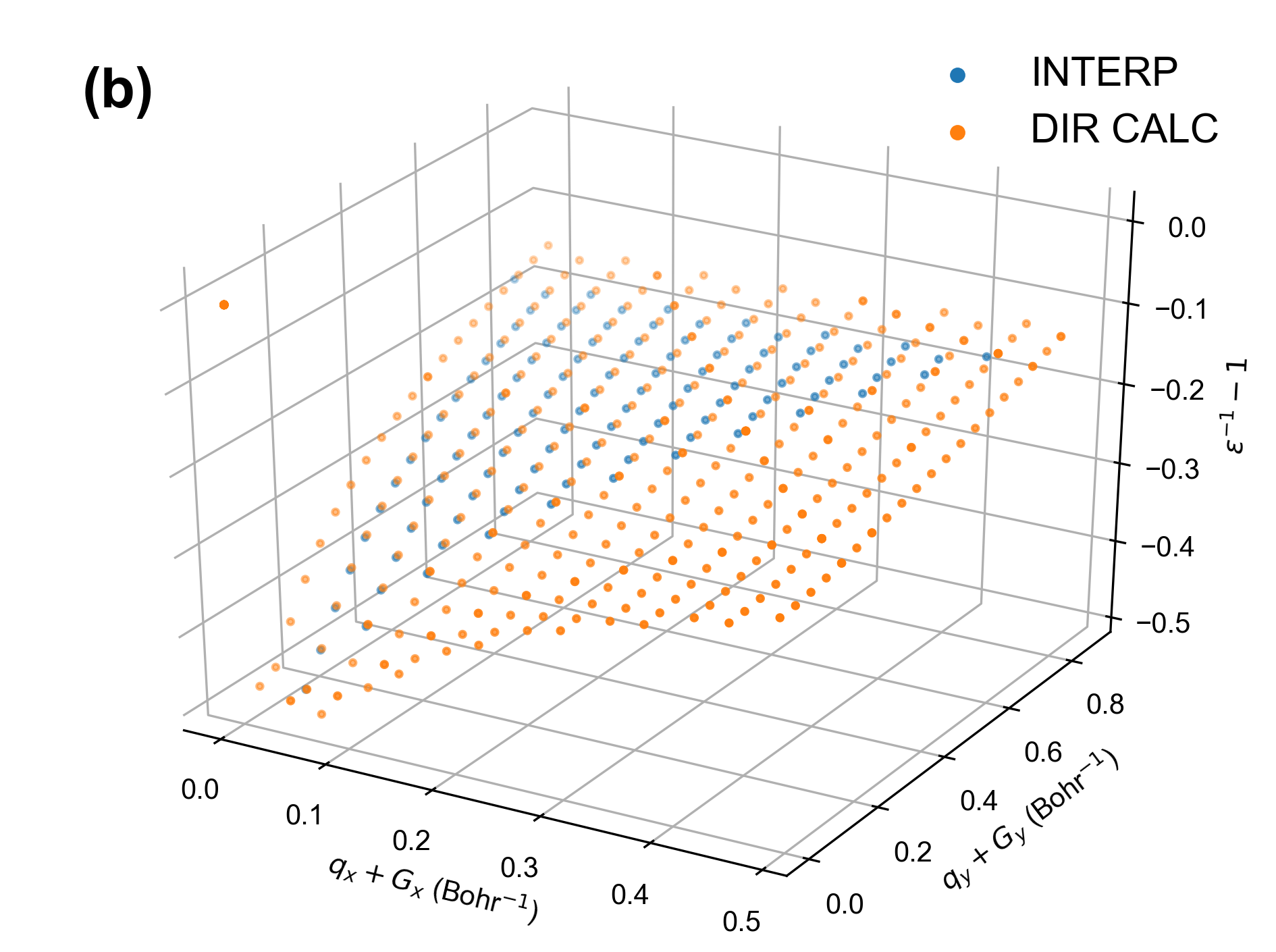}
    \centering
  \caption{\label{interp} 3D plot of in-plane diagonal elements of function $\epsilon^{-1}-\mathds{1}$ of SnS$_2$ substrate, with $\mathbf{G}$ vector subset $({G_z , G^{\prime}_z}) = ({0, 0})$ in a) full reciprocal space; b) a zoomed-in portion of a) that contains the irreducible Brillouin zone of the interpolated points (blue) in hBN $\mathbf{\Tilde{q}} + \mathbf{\Tilde{G}}$ subspace. 
  The $q_i + G_i , i=x, y$ are in-plane reciprocal space Cartesian coordinate in atomic unit (Bohr$^{-1}$). 
  The orange points are directly computed data points (``DIR CALC") in SnS$_2$ substrate $\mathbf{q} + \mathbf{G}$ subspace, while 
 the blue points are interpolated points (``INTERP") to (hBN) material $\mathbf{\Tilde{q}} + \mathbf{\Tilde{G}}$ subspace. Note that the orange points used for interpolating blue points in b) are beyond the first Brillouin zone of the substrate $\mathbf{q} + \mathbf{G}$ subspace. The single point at zero is the head element of $\epsilon^{-1}-\mathds{1}$, which is exactly zero for both material and substrate.}
\end{figure}

In this work we propose a general method for arbitrarily lattice-mismatched interfaces where it is not possible to map the $\mathbf{q} + \mathbf{G}$ vectors between the two subsystems. This approach applies a linear interpolation of the matrix elements on the substrate grid ($\mathbf{q} + \mathbf{G}$, $\mathbf{q'} + \mathbf{G'}$) to obtain their representation on the material grid ($\mathbf{\Tilde{q}} + \mathbf{\Tilde{G}}$, $\mathbf{\Tilde{q}'} + \mathbf{\Tilde{G}'}$), as shown in Figure~\ref{interp1}(b) and (c).
We note that we need to interpolate $\mathbf{q} + \mathbf{G}$ together between materials and substrates, which can completely remove the symmetry constrain. Interpolation of $\mathbf{q}$ only  as done in the past work~\cite{felipe2017nonuniform,kammerlander2012speeding}  will improve $\mathbf{q}$-sampling convergence speed but does not solve the periodicity or symmetry mismatch problem at interfaces. 
As this procedure requires a sampling of the $\mathbf{q}$ vectors over the full first Brillouin zone (FBZ), whenever necessary, the symmetry operators are used to reconstruct the grid in the FBZ from the grid in the irreducible Brillouin zone (IBZ).
Without loss of generality we choose the same size for vacuum in the $z$-direction for both subsystems; in this way the same out-of-plane reciprocal lattice ${G}_z$ components are obtained. 
In order to simplify the implementation, we neglect the in-plane off-diagonal elements of the substrate, i.e. we consider $v_C\chi^{s}_{\mathbf{G}, \mathbf{G}^{\prime}}(\mathbf{q}) \approx v_C\chi^{s}_{\mathbf{G} \mathbf{G}^{\prime}}(\mathbf{q}) \delta_{{G}_x , {G}_x^{\prime}} \delta_{{G}_y , {G}_y^{\prime}}$. As shown later for specific numerical examples (see Sec.~\ref{mismatch-interf-example}), this approximation works well in practice for mismatched 2D interfaces. 
For each set of matrix elements at fixed $\{{G}_z , {G}^{\prime}_z \}$, the standard bilinear interpolation technique~\cite{press2007numerical} is used to obtain the corresponding in-plane matrix elements $v_C\chi_{{G}_z, {G}^{\prime}_z}(\mathbf{\Tilde{q}}_{x,y} + \mathbf{\Tilde{G}}_{x,y})$ in the material subspace, interpolated from  $v_C\chi_{{G}_z, {G}^{\prime}_z}(\mathbf{q}_{x,y} + \mathbf{G}_{x,y})$ in  the substrate subspace. 
As shown in Figure~\ref{interp1}(b), the value of the response function $\epsilon^{-1}-\mathds{1}=v_C \chi$ at each $\Tilde{\mathbf{q}} + \Tilde{\mathbf{G}}$ point (denoted by the black cross overlaying the blue dots) is obtained by interpolating the values at the four nearest $\mathbf{q} + \mathbf{G}$ points (denoted by the red cross overlaying the orange dots). We note that the bilinear interpolation method can be applied only if all four nearest neighbours exist within the boundaries of $\mathbf{q} + \mathbf{G}$ space; otherwise the standard proximal interpolation method is applied, which considers only the nearest point on the grid (most likely at the boundary), as shown in Figure~\ref{interp1}(c). However, the values close to the boundary of $\mathbf{q} + \mathbf{G}$ space are very close to zero as shown in Figure~\ref{interp}(a).The bilinear interpolation method is fully general regardless of crystal symmetry, which can be applied to arbitrary interfaces.

By applying the interpolation method, we can obtain substrate $v_C\chi$ matrix elements at the material's $\mathbf{\Tilde{q}} + \mathbf{\Tilde{G}}$ grids, without any artificial strain~\cite{liu2019accelerating, ugeda2014giant, xuan2019quasiparticle}. As shown in Figure~\ref{interp}(a), the orange points are the $v_C\chi_{{G}_z, {G}^{\prime}_z}(\mathbf{q}_{x,y} + \mathbf{G}_{x,y})$ values computed at the substrate momentum space with full BZ, then we interpolate them to the blue points on the grids of material momentum space $v_C\chi_{{G}_z, {G}^{\prime}_z}(\mathbf{\Tilde{q}}_{x,y} + \mathbf{\Tilde{G}}_{x,y})$ (only elements in IBZ 
are shown here). The blue points fall smoothly on the surface of orange points which show a good interpolation quality. A zoomed-in picture is also shown in Figure~\ref{interp}(b). To show the generality of our method, we applied this interpolation method for hBN/phosphorene(BP)  interface, where BP has a rectangle lattice, sharply different from the hexgonal lattice hBN has (see SI Figure 1). We show again with our interpolation method, one can obtain the matrix elements of substrates at the material $\mathbf{\Tilde{q}} + \mathbf{\Tilde{G}}$ grids. Then we can compute the quasiparticle energies of this interface, at two systems' natural lattice constants, with the $\chi_{\text{eff}}$-sum method. 

\section{Computational details}

\subsection{Computational workflow}

The workflow of GW calculations for the interface is structured as follows. We first compute the reducible polarizabilities ($\chi$) for each subsystem separately and then we use them to obtain the effective polarizabilities ($\chi_{\text{eff}}^{m/s}$) using  Eq.~\ref{eq_chi_eff}. 
In case of lattice mismatch between the two subsystems, the matrix elements of the polarizability $\chi$ of the substrates are obtained on the same reciprocal space grid of the material (ML hBN in the practical applications of this work) by using the linear interpolation method described above. Next we sum them to obtain $\chi^{\text{tot}}$ (i.e. the $\chi_{\text{eff}}$-sum method).

Finally, in order to include the screening effect of the substrate on the material, the GW calculations are performed for the standalone hBN ML with the $\chi^{tot}$ obtained in the previous step. As we will discuss later, one can achieve further improvement for interfaces with strong hybridization by including corrections from ground state eigenvalues and wavefunctions of explicit interfaces.

\subsection{Numerical parameters}

In this work, we mainly focus on the quasiparticle energies of monolayer hBN/substrate interfaces as prototypical systems (where as substrates we will consider monolayer hBN itself and monolayer SnS$_2$). 
Density functional theory (DFT) ground state calculations based on the  Perdew-Burke-Ernzerhof (PBE) exchange-correlation functional~\cite{perdew1996generalized} have been performed using the open source plane-wave code Quantum ESPRESSO~\cite{Giannozzi_2017} with Optimized Norm-Conserving Vanderbilt (ONCV) pseudopotentials~\cite{hamann2013optimized} and a 80 Ry wave function cutoff. From structural relaxation we obtained lattice constants of 2.51 (\AA) and 3.70 (\AA) for the free-standing monolayer (ML) hBN and SnS$_2$, respectively.


GW calculations with the Godby-Needs plasmon-pole approximation~\cite{godby1989metal, oschlies1995gw} (PPA) were then performed using the Yambo~\cite{marini2009yambo} code. We chose PPA as a showcase for lower computational cost, but we can apply the same $\chi_{\text{eff}}$-sum and reciprocal-space interpolation method with full frequency integration as well without technical difficulty, with more computational cost. Importantly, to the best of our knowledge, only PPA models or static COHSEX approximation were used in past calculations for the dielectric screening of substrates  ~\cite{yan2011nonlocal, ugeda2014giant, bradley2015probing, qiu2017environmental, xuan2019quasiparticle, liu2019accelerating, naik2018substrate} and the obtained results were reasonably accurate. We used the same plasmon frequency for all calculations $\omega_p$=27.2 eV and found little variation of the results (i.e. within 20 meV, with $\omega_p$ from 24.5 to 30 eV).

The distance between the nearest periodic repetitions along the vacuum direction was set to be 20 Å. In order to speed up convergence with respect to vacuum sizes, a 2D Coulomb truncation technique was applied to dielectric matrices and GW self-energies~\cite{rozzi2006exact}. 
For bilayer hBN systems, we set the interlayer distance to the bulk value of 3.33(\AA) for both of the two different stacking configurations considered here ($AA^{\prime}$ and $AB$). The hBN/SnS$_2$ interlayer distance was set to 3.31 (\AA) as obtained from structural relaxation with vdW-corrected functionals~\cite{grimme2006semiempirical, barone2009role}. 

For each free-standing monolayer (``ML") unit cell, the GW self-energy cutoff is set to 15 Ry. The number of bands is set to 1000 (1500) for hBN (SnS$_2$) unit cell calculations. The exchange self-energy cutoff is set to 40 Ry. We use a $30 \times 30 \times 1$ ($20 \times 20 \times 1$) $\mathbf{k}$-points sampling for ML-hBN (ML-SnS$_2$) unit cell calculations, unless specified. 

GW calculations for the full explicit heterointerfaces have also been performed to
obtain ``exact'' reference results to benchmark the different methods for the substrate screening effects (see Sec.~\ref{method-compare}). 
The computational parameters for the full interface are set to keep consistency between supercells and unit cell calculations. 
Additional computational details and convergence tests can be found in SI \footnote{See Supplemental Material at [URL will be inserted by publisher], for additional details on examples of interpolation method for arbitrarily mismatched interface, numerical parameters and convergence of GW calculations~\cite{birowska2019impact, gao2016speeding, qiu2016screening, wu2017first, rozzi2006exact}.}.

\section{Results and discussions}
\subsection{Numerical comparison of different methods for substrate screening} \label{method-compare}

\begin{figure}
\includegraphics[width=0.40\textwidth]{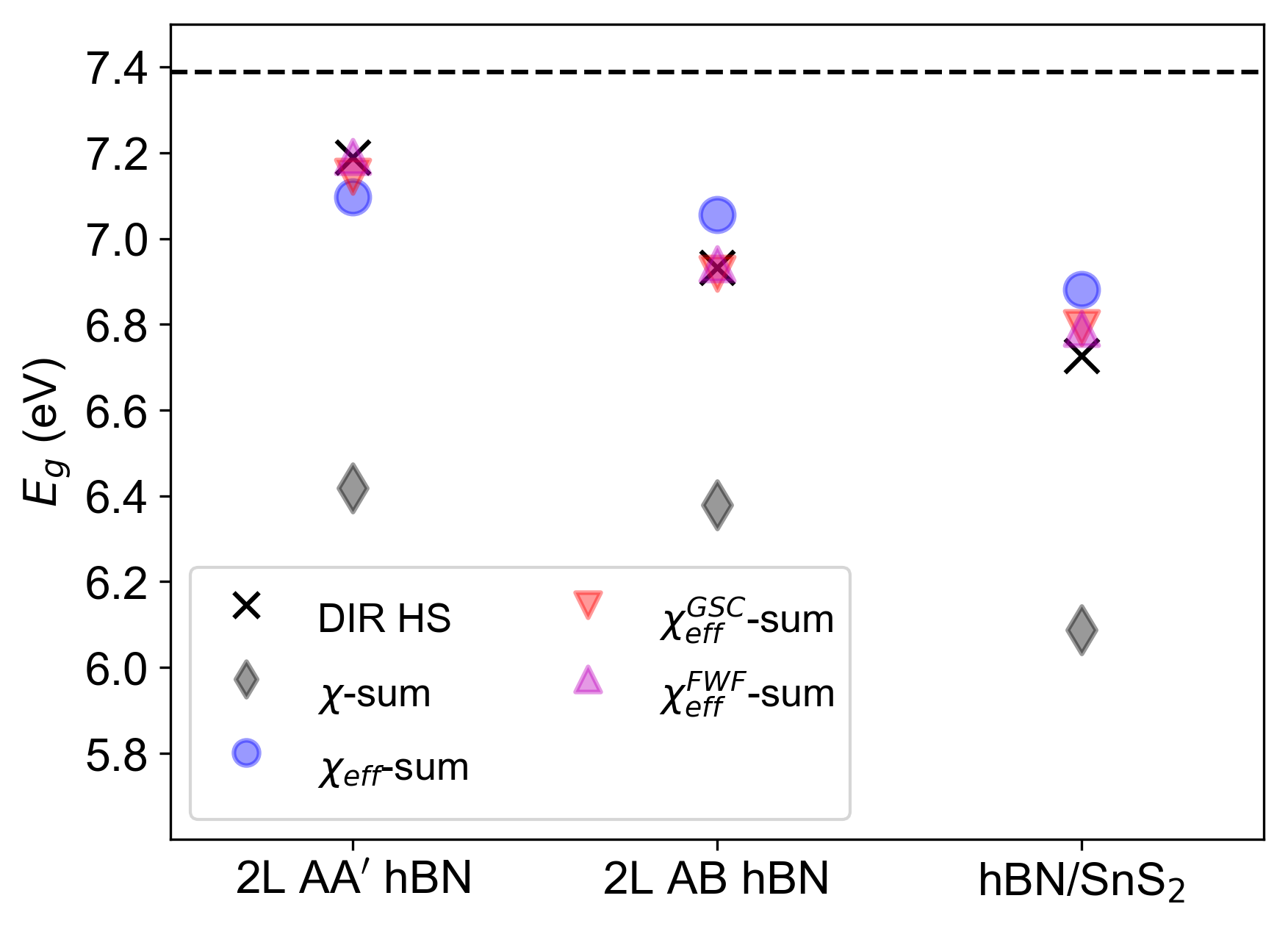}
\centering
 \caption{\label{figGW1} hBN direct band gap at $K$ from several interfaces with different approximations of substrate screening, compared with explicit interface calculations. The black dashed line is the direct band gap at $K$ of free-standing ML hBN. For each symbol in the figure, ``DIR HS" with black cross denotes direct GW calculation of explicit heterostructure; ``$\chi_{\text{eff}}$-sum" with blue circle denotes sum of effective polarizability approach by Eq.~\ref{eq_sum_chi_eff} with ground state inputs from free-standing ML hBN;  ``$\chi_{\text{eff}}^{GSC}$-sum" method with red down triangle denotes ``$\chi_{\text{eff}}$-sum" method with additional eigenvalue corrections from ground state interface eigenvalues (``GSC"); ``$\chi_{\text{eff}}^{FWF}$-sum" 
 with magenta up triangle denotes ``$\chi_{\text{eff}}$-sum" method with both ground state eigenvalues and wavefunctions from interfaces; ``$\chi$-sum" denotes non-interacting ``$\chi$-sum" method by Eq.~\ref{eq_sum_chi}.}
\end{figure}

After presenting in Sec.~\ref{method} with different approaches to approximate the total dielectric screening of an interface between two weakly interacting subsystems, in this section we discuss their accuracy in practical GW calculations. Results for
explicit interfaces will be used as a reference. Specifically, we computed the GW quasiparticle bandgaps of three interfaces: hBN/SnS$_2$, 2L-AB stacking hBN with two layers' atoms misaligned, and 2L-AA$^\prime$ stacking hBN interface with two layers' atoms aligned 
(the corresponding atomic structures are shown in Figure~\ref{atomic}). In order to keep the comparison of different methodologies as simple as possible, the calculations in this section are performed with fully commensurate interfaces, for both explicit and approximate interface calculations, as the results shown in Figure~\ref{figGW1}.   

From the explicit interface results in Figure~\ref{figGW1} we see that the direct band gap of hBN at the hBN/SnS$_2$ interface (black cross in the third column) is reduced by 0.8 eV compared with the isolated ML hBN (dashed line). This value is about four times of the band gap reduction for the bilayer hBN with respect to the isolated ML hBN (black cross in the first and second columns).
This is because ML SnS$_2$ has a much stronger dielectric screening ( $\epsilon_\infty \approx 17$) and a smaller electronic band gap ($\approx 2$ eV) compared to ML hBN, which has $\epsilon_\infty \approx 5$ and an electronic band gap of $\approx 7$ eV. This indicates the positive correlation between electronic band gap reduction and substrate dielectric screening, similar to previous discussions~\cite{cho2018environmentally, olsen2016simple, jiang2017scaling}.

Secondly, we find that the effective polarizability approach results
(``$\chi_{\text{eff}}$-sum" method, blue circle) are consistently in good agreement with the ones from explicit interface GW calculations (``Direct", black cross), i.e. within 0.2 eV. We improve the agreement by 50 meV with additional corrections from ground state eigenvalues of interfaces (``$\chi^{GSC}_{\text{eff}}$-sum" method, red triangle), which partially take into account the effect of interlayer couplings on eigenvalues at the DFT level. Moreover, by using interface ground state wavefunctions and eigenvalues (``FWF") as inputs for Green's function calculations, the results of the effective polarizability approximation (``$\chi^{FWF}_{\text{eff}}$-sum" method, green square) are further improved, i.e. with only 10 meV difference from the explicit interface GW calculations. While a similar approach was used in Ref.~\citenum{xuan2019quasiparticle}, the $\chi^{FWF}_{\text{eff}}$-sum method has a computational cost similar to that of the full interface GW calculation (although the evaluation of the dielectric matrix is more efficient) and is much more demanding than the other methods in Figure~\ref{figGW1} and Table~\ref{tab:table1}.
Therefore $\chi_{\text{eff}}$-sum and
$\chi^{GSC}_{\text{eff}}$-sum provide the best compromise between accuracy and computational cost. We note that for explicit hBN/SnS$_2$ interface, we had to apply 1.5\% strain to obtain commensurate supercells which may explain why this interface has slightly larger difference between $\chi_{\text{eff}}$-sum and explicit calculation than bilayer hBN. 

In sharp contrast to the methods discussed above, the non-interacting interlayer method based on Eq.~\ref{eq_sum_chi} (``$\chi$-sum" method, black diamond) gives results far from the explicit interface reference (e.g. with an error of about 0.6 eV). This indicates that the interlayer Coulomb interaction plays a dominant role in the electronic bandgap reduction by substrate screening.

\subsection{Diagonal approximation of substrate dielectric screening} \label{diag-compare}

In this section we will compare different diagonal approximations for the screening considering different numerical examples.
With ``in-plane  $\epsilon^{-1}$-diag'' we will denote an approximation that discards the in-plane off-diagonal elements of reducible polarizability $\chi$ in reciprocal space, i.e. $\chi_{\mathbf{G} \mathbf{G}^{\prime}}(\mathbf{q}, \omega) \delta_{{G}_x , {G}_x^{\prime}} \delta_{{G}_y , {G}_y^{\prime}}$. Similarly, ``out-of-plane $\epsilon^{-1}$-diag" will denote an approach that does not include the  out-of-plane off-diagonal elements of polarizability in reciprocal space, i.e. $\chi_{\mathbf{G} \mathbf{G}^{\prime}}(\mathbf{q}, \omega) \delta_{{G}_z , {G}_z^{\prime}}$. Analogous definitions will be used for $\epsilon$-diag.
\begin{figure}
    \includegraphics[width=0.40\textwidth]{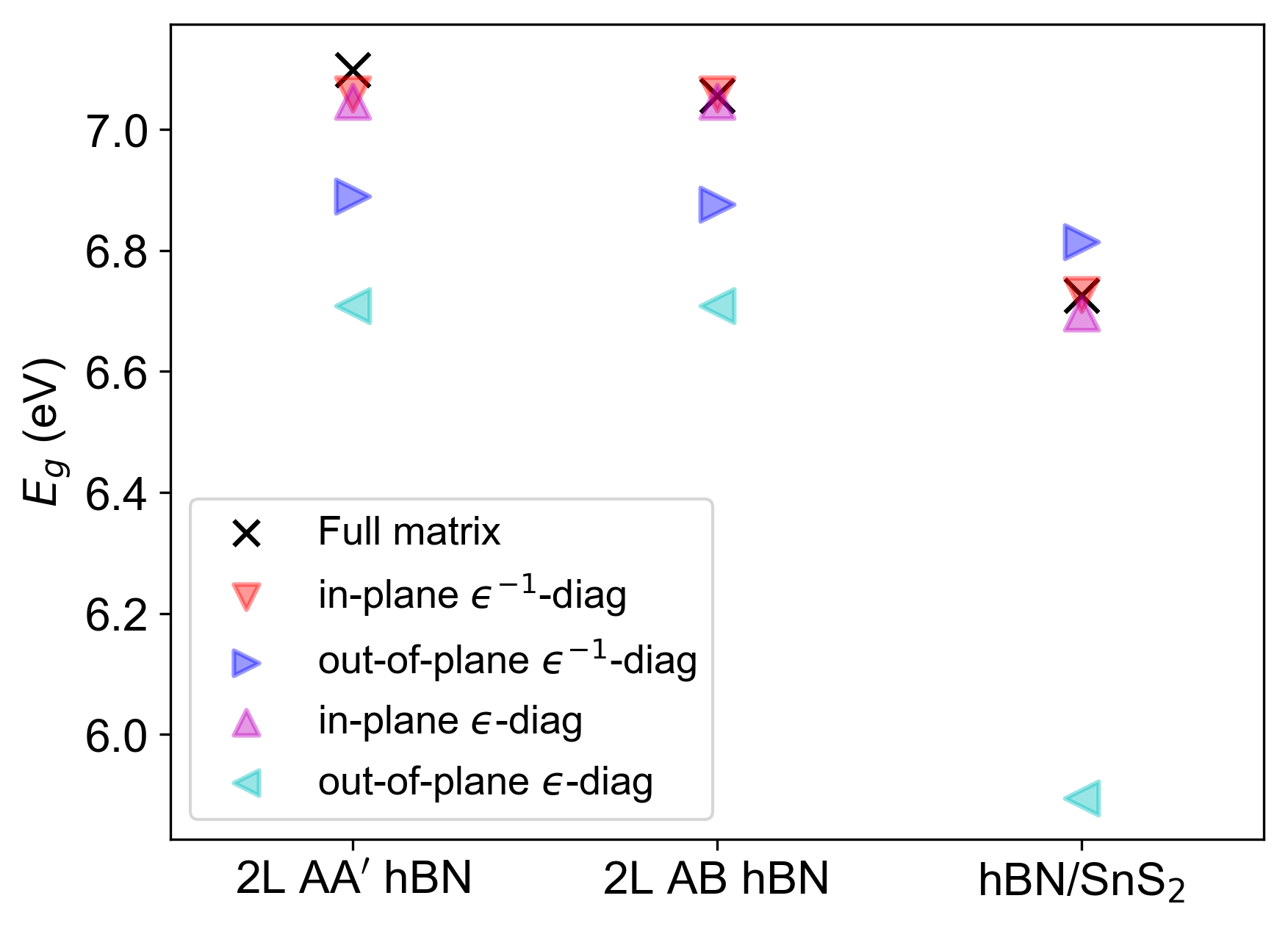}
    \centering
  \caption{\label{figGW2} GW results for hBN direct band gap at $K$  using ``$\chi_{\text{eff}}$-sum" or equivalently ``$\chi_0$-sum"  method to examine the effect of diagonal approximations.``Full matrix" in black cross denotes full dielectric matrices without any diagonal approximation as reference; ``in-plane  $\epsilon^{-1}$-diag" in red down triangle denotes diagonal approximation to in-plane elements of $\epsilon^{-1}$; ``out-of-plane $\epsilon^{-1}$-diag" in dark blue right triangle denotes diagonal approximation to out-of-plane elements for $\epsilon^{-1}$; ``in-plane  $\epsilon$-diag" in magenta up triangle denotes diagonal approximation to in-plane elements of $\epsilon$; ``out-of-plane $\epsilon$-diag" in light blue left triangle denotes diagonal approximation to out-of-plane elements of $\epsilon$.}
\end{figure}

The GW quasiparticle gaps with different diagonal approximations for the hBN bilayer in two different confomations (AA$^\prime$/ AB) and the hBN/SnS$_2$ interface are shown in Figure~\ref{figGW2}. We find that for both the $\epsilon^{-1}$ and $\epsilon$ diagonal approximations, neglecting out-of-plane off-diagonal elements of the substrate (``out-of-plane $\epsilon^{-1}$-diag" and ``out-of-plane $\epsilon$-diag", dennoted by dark blue right triangle and light blue left triangle, respectively) causes a large discrepancy of the bandgaps (i.e. from 0.2 to 0.8 eV) with respect to the ``exact'' result obtained from the full screening matrix (``Full matrix", black cross). In contrast, the results obtained by neglecting in-plane off diagonal elements (in-plane $\epsilon^{-1}$/ in-plane $\epsilon$-diag, red down triangle/ magenta up triangle) are similar to those with the full screening matrix with deviations within 50 meV. 
This means the inhomogeneity effect of out-of-plane substrate screening on quasiparticle energies is much stronger than the one of in-plane  substrate screening, because the out-of-plane direction is along the non-periodic (vacuum) direction with dramatically inhomogeneous charge distribution, compared to the in-plane periodic direction.

Besides the overall difference of diagonal approximation along different directions, we also distinguish the difference between $\epsilon^{-1}$-diag and $\epsilon$-diag approach in each case. 1) Along the in-plane direction, the difference between different approaches is negligible, i.e. less than 10 meV. 2) Along the out-of-plane direction, the out-of-plane $\epsilon^{-1}$-diag results (dark blue right triangle) are much closer to the full dielectric matrix results (black cross) than the out-of-plane $\epsilon$-diag results (light blue left triangle) in Figure~\ref{figGW2}.  
This is consistent with our earlier speculation that the $\epsilon^{-1}$-diag may be a better (weaker) approximation, because the off-diagonal elements of irreducible polarizability $\chi_0$ contribute to $\chi$ or $\epsilon^{-1}$ during its matrix inversion, which is completely missing in the $\epsilon$-diag approximation. 

Moreover, the in-plane inhomogeneity is relatively larger when there is stronger interlayer coupling with atoms aligned perfectly for chemical bonding. For example, the in-plane inhomogeneity of bilayer hBN with atoms aligned (e.g. 2L AA$^{\prime}$ hBN in Figure~\ref{atomic} (b); both in-plane $\epsilon^{-1}$-diag (red down triangle) and in-plane $\epsilon$-diag (magenta up triangle) results have 40 meV difference from the ``Full matrix" results in the first column of Figure~\ref{figGW2}), is larger than the interfaces with atoms misaligned (e.g. 2L AB hBN and hBN/SnS$_2$ hectorstructure in Figure~\ref{atomic} (a) and (c); both in-plane $\epsilon^{-1}$-diag and in-plane $\epsilon$-diag results have no difference from ``Full matrix" results in the second and third columns of Figure~\ref{figGW2}).

\begin{figure}
    \includegraphics[width=0.40\textwidth]{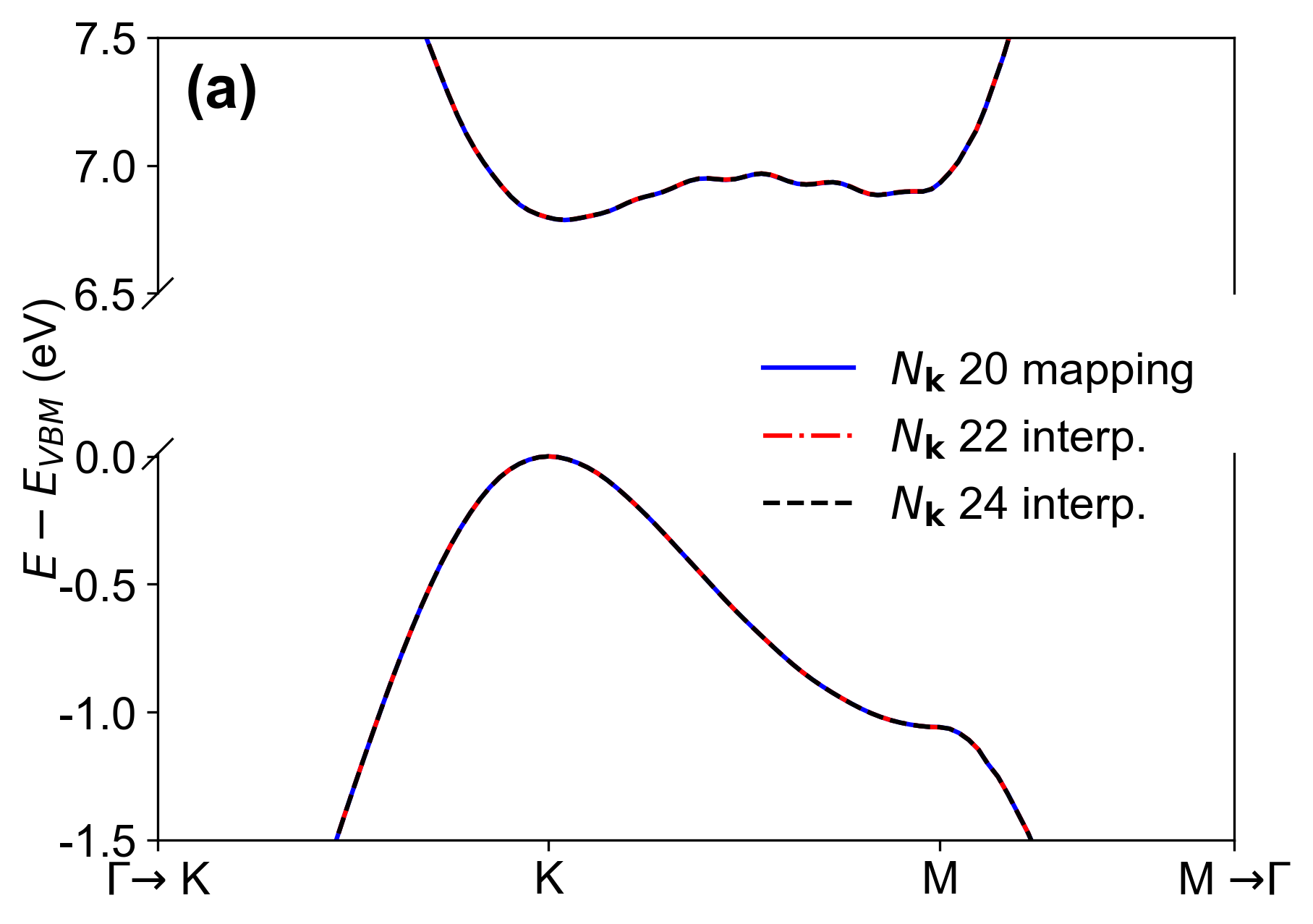}\\
    \includegraphics[width=0.40\textwidth]{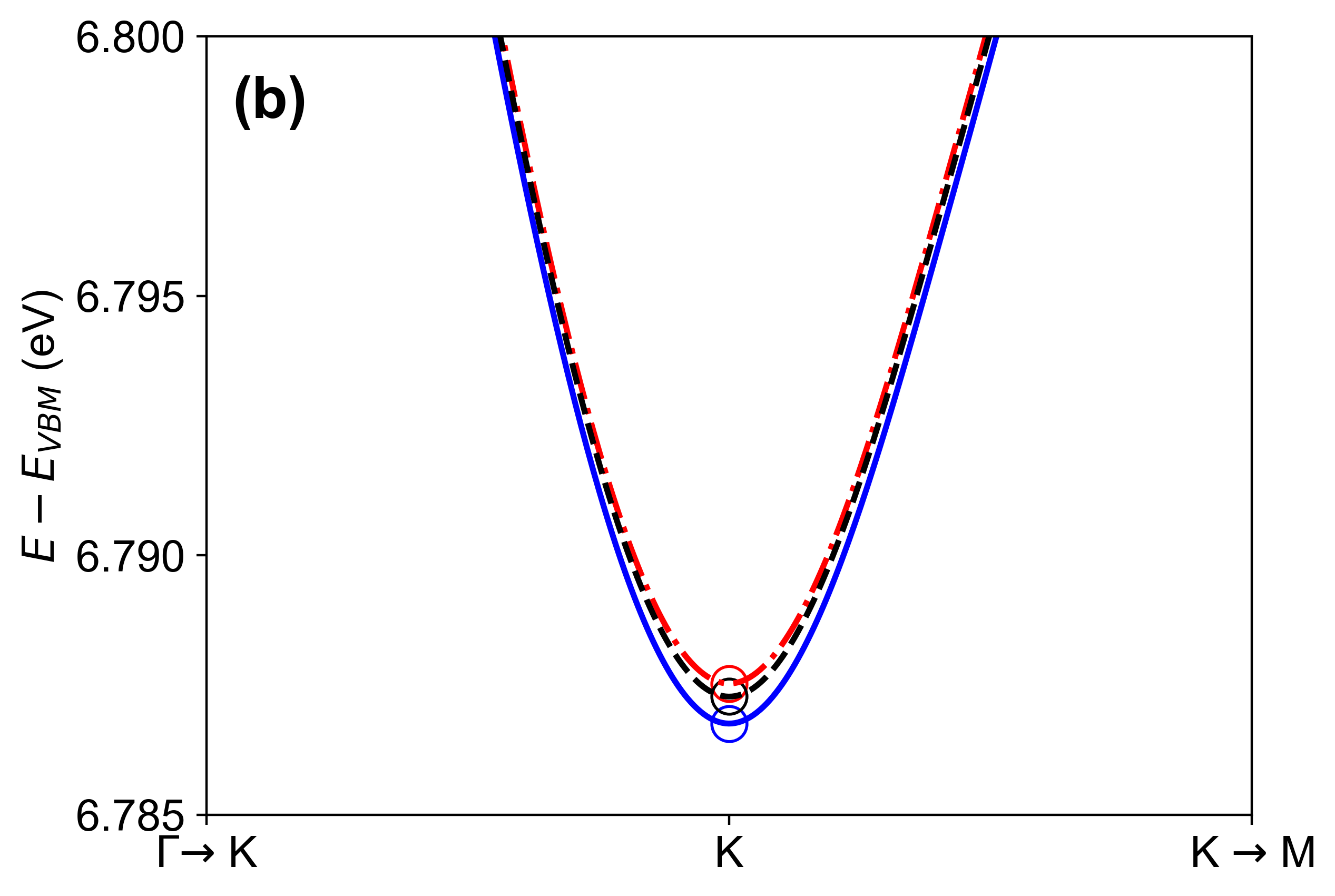}
    \centering
  \caption{\label{interp-match} GW band structure of hBN at hBN/ stretched SnS$_2$ interface, referenced to valence band maximum (VBM) by using ``$\chi_{\text{eff}}$-sum" method. The blue solid line is computed with commensurate $\mathbf{q}$-point sampling with the reciprocal space mapping approach, while the red/black dash line results are computed with  incommensurate $\mathbf{q}$-point sampling with the reciprocal-space linear-interpolation approach. a) shows both valence and conduction band edges; b) shows only the conduction band edge close to $K$.}
\end{figure}

\subsection{Lattice mismatched hBN/SnS$_2$ interface}\label{mismatch-interf-example}

In order to benchmark the reciprocal-space linear-interpolation method introduced in Sec.~\ref{RLM} 
and Table~\ref{tab:table1}, we consider the hBN/SnS$_2$ interface. A strain of 1.5\% was applied to SnS$_2$ to match the hBN lattice constant with a 2:3 ratio in each direction of the plane (namely $2L^{\text{SnS}_2}=3L^\text{hBN}$). By using a commensurate $\mathbf{q}$-point sampling for the two subsystems with a 2:3 ratio, a mapping of the 
$\mathbf{q} + \mathbf{G}$ vectors is possible and traditional methods for the substrate effect can be applied to produce a reference results for our new interpolation method (which, instead, will be used with an incommensurate $\mathbf{q}$-point sampling). 
We computed the GW band edges near the high symmetry point $K$ of hBN on the SnS$_2$ substrate with the $\chi_{\text{eff}}$-sum method at different $\mathbf{k}$-point samplings, as shown in Figure~\ref{interp-match}. The mesh for $\mathbf{q}$-point sampling was chosen to be identical to the $\mathbf{k}$-point sampling. Specifically, the reference calculations  were performed with  the hBN unit cell calculation with $20 \times 20 \times 1$ and $30 \times 30 \times 1$ $\mathbf{k}$-point sampling for the units cells of SnS$_2$ and hBN, respectively (this choice satisfies the 2:3 ratio for each inplane direction). The reference result obtained from the $\mathbf{q} + \mathbf{G}$ mapping is shown in Figure~\ref{interp-match} (blue curve labelled by ``$N_\mathbf{k}$  20 mapping").
To apply our interpolation technique, it is not necessary to use commensurate grids and we compare instead the results for two different choices of the $\mathbf{q}$-point sampling for SnS$_2$. Specifically, in Figure~\ref{interp-match} we show the results for the
$22 \times 22 \times 1$ (red dashed line, ``$N_\mathbf{k}$ 22 interp.") and $24 \times 24 \times 1$ (black dashed line, ``$N_\mathbf{k}$ 24 interp.") $\mathbf{q}$-point grids, which do not allow for a mapping of the reciprocal space vectors and would be impossible to treat without our interpolation method. 
The results in Figure~\ref{interp-match}(a) show that the GW band structure with interpolation (red and black dashed lines) is nearly identical to the one based on the mapping (blue solid line), with differences  smaller than 1 meV (as can be seen by zooming-in the conduction band edge in Figure~\ref{interp-match}(b)). This comparison demonstrates the excellent numerical accuracy of our linear interpolation method, which could have also been expected from the high quality of the interpolation in Figure~\ref{interp}.

\begin{figure}
    \includegraphics[width=0.40\textwidth]{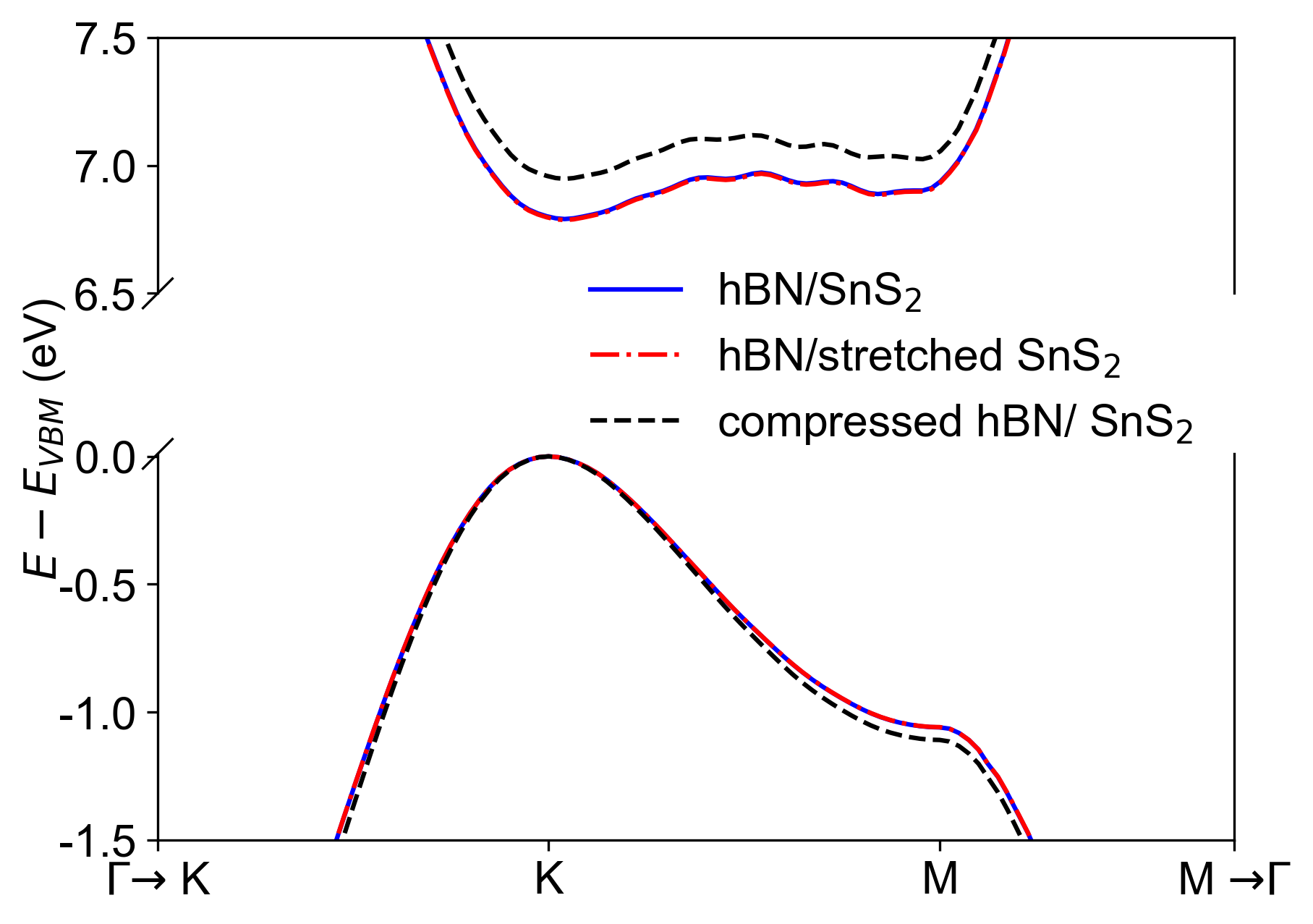}
    \centering
  \caption{\label{interp-mismatch} GW band structures of hBN with stretched SnS$_2$ substrate (``hBN/stretched SnS$_2$", red dash-dotted line), hBN with SnS$_2$ substrate with no strain (``hBN/SnS$_2$", blue solid line), compressed hBN with SnS$_2$ substrate (``compressed hBN/SnS$_2$", black dashed line), respectively.}
\end{figure}

Finally, we use our new interpolation method to better understand the effect of the strain on quasiparticle energies.  In Figure~\ref{interp-mismatch}, the blue curve corresponds to the
the hBN bandstructure on the SnS$_2$ substrate without strain for either system as obtained from the interpolation scheme described above. 
These results are compared with those obtained for the interface by applying strain either to compress the hBN lattice parameter or to stretch SnS$_2$. We found that even a 1.5$\%$ compressive strain for hBN (black dashed line), the conduction band edge changes by 0.2 eV. Since we are focusing on the band structure of hBN states, the application of the strain to the SnS$_2$ substrate leads to negligible changes (red dash-dotted line). This result highlights the high sensitivity of quasiparticle band structures to strain. 

We note that for a proof of principle and benchmark purpose, we chose systems with similar crystal symmetry, i.e. hexagonal lattice, in this work. However, our interpolation method can be applied to general interfaces with very different crystal symmetry, e.g. interface between hexagonal and rectangle lattices, as the example of hBN/phosphorene interface shown in SI Figure 1. This is not possible by using the previous $\mathbf{q}+\mathbf{G}$ mapping approach.
Our reciprocal-space linear-interpolation method makes possible the GW calculations of interfaces composed by two materials with very different lattice parameters and symmetry, at the cost of primitive cell calculations only.

\section{Conclusion}

In this work, we theoretically and numerically examined the existing methods to approximate substrate dielectric screening effect on quasiparticle energies, through hBN heterostructures as prototypical examples. We clarified the theoretical equivalence between the sum of effective reducible polarizability approach ($\chi_{\text{eff}}$-sum) and sum of irreducible polarizability of interface systems ($\chi_{0}$-sum), at the RPA level. We numerically compared the GW calculations of 2D interfaces with several approximations, and found excellent agreements between $\chi_{\text{eff}}$-sum and the explicit interface calculations. Further improvement can be achieved by including the ground state corrections of eigenvalues (and wavefunctions) from explicit interfaces.  We further evaluated the importance of non-diagonal elements of $\epsilon$ and $\epsilon^{-1}$ from substrates on quasiparticle energies of 2D interface. 
Most importantly, we developed an accurate reciprocal-space linear-interpolation technique for arbitrarily lattice-mismatched interfaces, which can be used to compute the interface polarizability for GW quasiparticle energies without any artificial strain, at the cost of only primitive cell calculations.

\section*{ACKNOWLEDGMENTS}

We thank Diana Qiu and Felipe H. da Jornada for helpful discussions. This work is supported by the National Science Foundation under grant number DMR-1760260. This research used resources of the Scientific Data and Computing center, a component of the Computational Science Initiative, at Brookhaven National Laboratory under Contract No. DE-SC0012704, the lux supercomputer at UC Santa Cruz, funded by NSF MRI grant AST 1828315, the National Energy Research Scientific Computing Center (NERSC) a U.S. Department of Energy Office of Science User Facility operated under Contract No. DE-AC02-05CH11231, the Extreme Science and Engineering Discovery Environment (XSEDE) which is supported by National Science Foundation Grant No. ACI-1548562 \cite{xsede}.

\bibliographystyle{apsrev4-1}
\bibliography{ref}

\end{document}